\documentclass{iaus}
\usepackage{graphicx}
\usepackage{natbib}

\def\sss{\scriptscriptstyle}
\def\U{{\sss \!U}}

\def\nuU{\nu_\U}

\title[Troubles with the spin] 
{QPOs in microquasars: the spin problem}

\author[Gabriel T\"{o}r\"{o}k et al.]   
{%
  Gabriel T\"{o}r\"{o}k$^{1,3}$, Marek A. Abramowicz$^{1,2,3}$, \break Zden\v{e}k Stuchl\'{\i}k$^{1}$ \and Eva \v{S}r\'{a}mkov\'{a}$^{1,3}$}

\affiliation{$^1$Institute of Physics, Faculty of Philosophy and Science, Silesian University in Opava, Bezru\v{c}ovo n\'{a}m. 13,  CZ-74601 Opava, Czech Republic. \break email: terek@volny.cz, zdenek.stuchlik@fpf.slu.cz, sram\_eva@centrum.cz\\[\affilskip]
$^2$Department of Physics, G\"{o}teborg University, S-412~96 G\"{o}teborg, Sweden.\break email: marek@fy.chalmers.se\\[\affilskip]
$^3$Copernicus Astronomical Centre, Polish Academy of Sciences, PL-00~716 Warszawa, Bartycka~18, Poland}

\pubyear{2006}
\volume{240}  
\pagerange{119--121}
\date{10th October and in revised form ??}
\jname{Proceedings of XXVIth IAU General Assembly 2006}
\editors{B. Hartkopf, E. Guinan \& P. Harmanec, eds.}
\begin{document}

\maketitle

\begin{abstract}

In the Galactic microquasars with double peak kHz quasi-periodic
oscillations (QPOs), the ratio of the two frequencies is 3:2. This
supports the suggestion that double peak kHz QPOs are due to a non-linear resonance
between two modes of accretion disk oscillations. For the microquasars with known mass, we briefly compare the black hole spin estimates based on the orbital resonance model with the recently reported spin predictions obtained by fitting the spectral continua. Results of these two approaches are not in good agreement. We stress that if the spectral fit estimates are accurate and can be taken as referential (which is still questionable), the disagreement between the predicted and referential values would represent a rather generic problem for any relativistic QPO model, as no spin influence would appear in the observed $1/M$ scaling of the QPO frequencies. The epicyclic frequencies relevant in these models are often considered to be equal to those of a test particle motion. However modifications of the frequen\-cies due to the disc pressure
or other non-geodesic effects may play an important role, and the inaccuracy introduced in the spin estimates by the test particle approximation could be crucial.
\end{abstract}

\firstsection 
\section{Estimating the black hole spin from the resonance models}

The resonance model \citep[][]{KluzniakAbramowicz2000}  explains the twin peak QPOs as being caused by a 3:2 non-linear resonance between two global modes of oscillations in accretion flow in strong gravity. The modes in resonance are often assumed to be the epicyclic modes. The \emph{orbital resonance model} \citep[see][]{ KluAbr:2003:aph} demonstrates that {\it fluid accretion flows} admit two linear quasi-incompressible modes of oscillations, radial and vertical, with corresponding eigenfrequencies equal to the radial and vertical epicyclic frequencies for free particles \citep{Ali-Gal:1981:GENRG2, Now-Leh:1998:TheoryBlackHoleAccretionDisks:}. According to the resonance hypothesis, the two modes in resonance have eigenfrequencies $\nu_{\rm r}$ (radial epicyclic frequency) and $\nu_{\rm v}$ (vertical epicyclic frequency $\nu_{\theta}$ or Keplerian frequency $\nu_{\rm K}$). Several resonances of this kind are possible and have been  discussed (see, e.g., \citealt[]{AbrKlu:2004}).

Formulae for the Keplerian $\nu_{\mathrm{K}}$ and the epicyclic frequencies $\nu_{\rm r}$ and $\nu_{\theta}$  in the field of a Kerr black hole with mass $M$ and spin $a$ are well known, and have the general form
\begin{equation}
\label{eq:oneoverM:theory}
\nu=\left ({{GM_0}\over {r_G^{~3}}}\right )^{1/2}f_\mathrm{i}(x,\,a)~~\doteq 32.3\left(\frac{M_0}{M_\odot}\right)f_\mathrm{i}(x,\,a)\,\mathrm{kHz},\quad\mathrm{i}\in{\mathrm{K},~\mathrm{r},\theta}
\end{equation}
where $f_\mathrm{i}(x,a)$ are functions of a dimensionless black hole spin $a$ and a dimensioless radial coordinate \mbox{$x\!=r/M$}.
For a $n\!:\!m$ orbital resonance, the dimensionless resonance radius $x_{\mathrm{n}:\mathrm{m}}$ is determined as a function of spin $a$ by an equation 
$
\mathrm{n}\nu_{\rm r}\!= \mathrm{m}\nu_{\rm v}~(\nu_\mathrm{v}\!=\nu_\theta\,~\mathrm{or}~\,\nu_\mathrm{K})
$
\footnote{Because of the properties of Kerr black hole spacetimes, \emph{any} relativistic model of black hole QPOs should be rather sensitive to the spin $a$, however this sensitivity can be negligible on large scales of mass (\citealt[]{Abr-etal:2004:apj}).}. 
Thus, from the observed frequencies and from the estimated mass, one can determine the relevant spin (\citealt[]{AbramowiczKluzniak2001,TAKS}). We summarize the spin estimates for the three microquasars in Table \ref{table:1}.

\newlength{\sirkaA}\newlength{\sirkaB}\settowidth{\sirkaA}{+}\settowidth{\sirkaB}{-}\advance\sirkaA by -\sirkaB
\newcommand{\csp}{$\hspace{\sirkaA}$}
\begin{table*}[t!]
\label{table:1}
\begin{center}      
\caption[]{\label{Table3} Spin estimates from the resonance models for microquasars \citep[for details and other considered resonances see, e.g.,][]{TAKS, Tor:2005:ASN}. Spin intervals correspond to the 1$\sigma$ uncertainty in mass, the small error resulting from uncertainty of the frequency measurement is for XTE 1550-564 (GRO 1655-40, GRS 1915+105) up to $0.03$ (0.01, 0.05).}
\begin{tabular}{ l  l  l  l  l }    
\hline
~& \multicolumn{4}{c}{{{Interval of possible spin $a$ relevant for}}}\\
Model for &~~~~~1550--564 &~~~~~1655--40 &~~~~~1655--40$^*$ &~~~~~1915+105\\
\hline
\hline\\
3:2 [$\nu_{\theta},~\nu_r$] ~~&
+0.89~---~+0.99 &+0.96~---~+0.99 &+0.88~---~+0.93 &+0.69~---~+0.99\\  
\hline
2:1 [$\nu_{\theta},~\nu_r$] ~~&+0.12~---~+0.42 &+0.31~---~+0.42 &+0.10~---~+0.25 & \csp -0.41~---~+0.44\\ 
\hline
3:1 [$\nu_{\theta},~\nu_r$] ~~& +0.32~---~+0.59 & +0.50~---~+0.59 &+0.31~---~+0.44 & \csp-0.15~---~+0.61\\ 
\noalign{\smallskip}
\hline
     \end{tabular}
     \end{center}
$^{*}$ The two columns for GRO 1655--40 indicate numbers following from the two different mass analysis - \cite{Beer2002} vs. \cite{Greene2001}. Note that while the spin estimates from the 3:2 parametric resonance is for both the cases similar ($a\approx 0.9$), for the other models, the given mass range implies a large range of spins.
\end{table*}

\vspace{-.5cm}

\section{Comparison with the fits of spectral continua}

Except for one case, all the resonances considered in \cite{TAKS} are consistent with reasonable values of the black hole spin covering the range $a\in(0,~1)$. In particular, the 3:2 epicyclic parametric (internal) resonance model, supposed to be the most natural one in Einstein gravity \citep{Hor:2005:ASN:}, implies the spin $a\sim 0.9$.

The most recent results of the spectral fits correspond for GRO~1655--40~ to the spin $a\in(0.65,~0.75)$, and for GRS~1915+105 to $a\!>0.98$ \citep[][]{McC-etal:2006:APJ:}. Obviously, the value for GRS~1915+105 is in agreement with the prediction of the 3:2 parametric epicyclic resonance model, but the same prediction for GRO~1655--40 does not match the spectral fitting. No particular resonance model considered so far can cover the spectral limits to the spin for both microquasars. It could be interesting that the recently proposed 3:2 periastron precession resonance \citep{Bur:2005:RAG:} implies  the spin of GRO~1655--40 to be $a\sim 0.7$. Nevertheless, eventuall periastron precession resonance requires the spin of GRS~1915+105 $a<0.8$ which is in strong disagreement with the spectral fitting limit, $a\!>0.98$.

\vspace{-.5cm}

\section{Troubles with the spin: $1/M$ scaling}
In principle one cannot exlude the possibility of different mechanisms exciting the high frequency QPOs in different sources, but there are many indicies that the mechanism is the same or similar \citep[e.g.,][]{Kli:2005:ASN:, Tor-etal:06}.
\citet{McClintockRemillard2003} found that the upper QPO frequency in microquasars scales well as
$\nuU\! =\!2.793 ( {M_0/M_{\odot}})^{-1}\, \mathrm{kHz}$ which is in good agreement with the 1$/M$ scaling of the first term in equation (\ref{eq:oneoverM:theory}).

On the other hand the exact 1$/M$ scaling holds only for the fixed value of the spin $a$ as functions $f_\mathrm{i}$ in equation (\ref{eq:oneoverM:theory}) are sensitive to the spin. The spectral limits to the spin for the two microquasars are \emph{very different}: $a\!\sim0.7$ vs. $a\!>0.98$, and, in addition, functions $f_\mathrm{i}(a,x)$ are more sensitive to the value of the spin when it is close to $a\!=1$ \citep[e.g.,][]{tor-stu:05:AA}. Hence, if the spin values obtained from the spectral fits are correct, the observed high frequency QPOs do not show sensitivity to the spin $a$ under the assumption of a unified QPO model. This is a serious problem for any relativistic QPO model handling with the orbital and epicyclic frequencies  (\ref{eq:oneoverM:theory}).

\vspace{-.5cm}

\section{Requirement of a more realistic description}

It was found recently that the pressure effects may have a strong influence on the oscillation frequencies. \cite{Sra:2005:ASN:} and \cite{Bla-etal:06} studied properties of the radial ad vertical epicyclic modes of slightly non-slender tori within Newtonian theory using the Paczy\'nski-Wiita potential, and found the epicyclic frequencies to decrease with increasing thickness of the torus. The same behaviour was found for the resonant radius where the frequencies are in a 3:2 ratio, which on the contrary implies {\it increase} of the resonant frequencies. Considering the appropriate corrections to frequencies in the Kerr metric, one can reestimate the values of the spin using the resonance model. If the results in the Kerr metric were following the same trend as those in the Paczy\'nski-Wiita case, the spin for some configurations can be {\it lower} than previously estimated.
\footnote{In case of the 3:2 parametric resonance, the maximal realistic increase of the resonant frequency due to the pressure effects is about 15 percent \citep{Bla-etal:06}, which for GRO 1655-40 and the mass estimate by Beer \& Podsiadlowski would lower the spin down to $a\sim$0.8.}

\begin{acknowledgments}
This research is supported by the Czech grant MSM~4781305903. 
\end{acknowledgments}

\end{document}